\documentclass[10pt]{article}
\usepackage{graphicx}
\usepackage{amssymb}

\def\Title#1{\begin{center} {\Large #1 } \end{center}}
\def\Author#1{\begin{center}{ \sc #1} \end{center}}
\def\Address#1{\begin{center}{ \it #1} \end{center}}

\newcommand\pubblock{\rightline{\begin{tabular}{l} Proceedings of the Second Annual LHCP\\ \pubnumber\\
         \pubdate  \end{tabular}}}

\newenvironment{Abstract}{\begin{quotation} \begin{center} 
             \large ABSTRACT \end{center}\bigskip 
      \begin{center}\begin{large}}{\end{large}\end{center} \end{quotation}}

\newenvironment{Presented}{\begin{quotation} \begin{center} 
             PRESENTED AT\end{center}\bigskip 
      \begin{center}\begin{large}}{\end{large}\end{center} \end{quotation}}

\def\beq{\begin{equation}}
\def\eeq#1{\label{#1}\end{equation}}
\def\eeqn{\end{equation}}

\def\beqa{\begin{eqnarray}}
\def\eeqa#1{\label{#1}\end{eqnarray}}
\def\eeqan{\end{eqnarray}}

\let\bar=\overbar

\def\Dslash{\not{\hbox{\kern-4pt $D$}}}
\def\dslash{\not{\hbox{\kern-2pt $\del$}}}

\def\msb{{\bar{\ssstyle M \kern -1pt S}}}

\usepackage{color}
\usepackage{floatflt}

\newcommand\pubnumber{ CMS CR-2014/225 }

\newcommand\pubdate{\today}

\def\affiliation{
On behalf of the CMS Experiment, \\
The Rockefeller University \\
1230 York Avenue, New York, NY 10065, U.S.A }

\begin{document}

\large
\begin{titlepage}
\pubblock

\vfill
\Title{  DIFFRACTION, FORWARD PHYSICS AND SOFT QCD RESULTS FROM CMS  }
\vfill

\Author{ ROBERT CIESIELSKI }
\Address{\affiliation}
\vfill
\begin{Abstract}

We report on the recent CMS measurements of soft hadron-hadron production, including the measurements of inclusive single- and double- diffractive cross sections, as well as the measurement of pseudorapidity distributions, and of the leading charged particle and leading jet cross sections. We present the cross sections for low-$p_\mathrm{T}$ forward jets production, and production of dijets with large rapidity separation (central-forward, and forward-backward jets). Studies of double parton scattering using four-jet and W$+$2-jet events, and the cross sections for Drell-Yan with associated jet production are discussed as well. The results, corresponding to the proton-proton center-of-mass energy of 7 or 8 TeV, are compared to predictions of various Monte Carlo models. 

\end{Abstract}
\vfill

\begin{Presented}
The Second Annual Conference\\
 on Large Hadron Collider Physics \\
Columbia University, New York, U.S.A \\ 
June 2-7, 2014
\end{Presented}
\vfill
\end{titlepage}
\def\thefootnote{\fnsymbol{footnote}}
\setcounter{footnote}{0}
%

\normalsize 


\section{Introduction}

Measurements of particle yields and kinematic distributions at the LHC contribute to a better understanding of the mechanisms of hadron production, and especially the relative roles of soft and hard scattering. Most of the particles produced in proton-proton collisions arise from semi-hard (multi)parton scatterings and/or diffractive processes, which are modeled phenomenologically; hence experimental results provide important input for tuning various models and event generators. When a hard scale is present in the process (e.g. when the final state includes high-pT jets, W or Z bosons, etc.) perturbative QCD (pQCD) is applicable and the dynamics can be formulated in terms of partons. The data are usually successfully described by pQCD calculations within the framework of collinear factorization and Dokshitzer-Gribov-Lipatov-Altarelli-Parisi (DGLAP) evolution equations. At the LHC, evidence for alternative approaches like the Balitski-Fadin-Kuraev-Lipatov (BFKL) approximation can be searched for, by studying dijet events widely separated in rapidity. Also, access to the region of low proton longitudinal momentum fractions x carried by the parton, where the parton densities are large, allows to study double parton scattering (DPS), which consists of two simultaneous hard interactions in the same collision. In this paper, we review the recent CMS results on diffraction, soft QCD and forward physics, and discuss their comparison to various theoretical predictions.

\section{Results}

\begin{figure}[b]
\centering
\includegraphics[height=1.61in]{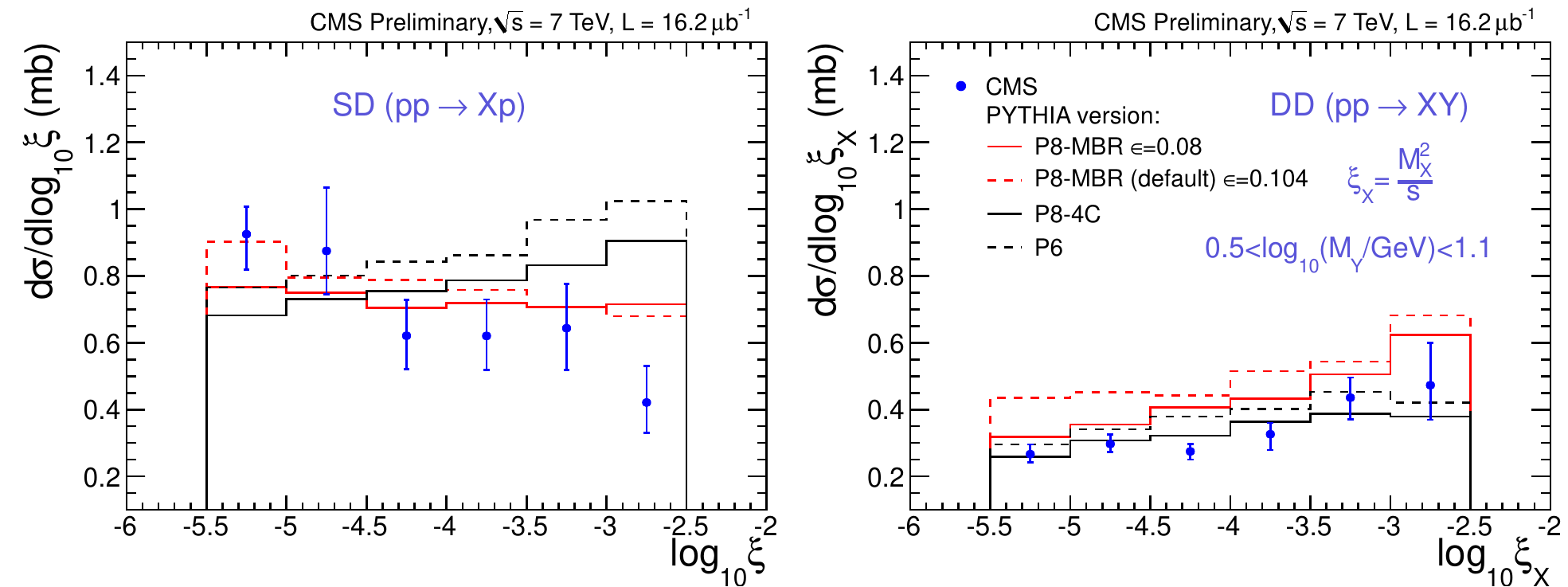}
\includegraphics[height=1.61in]{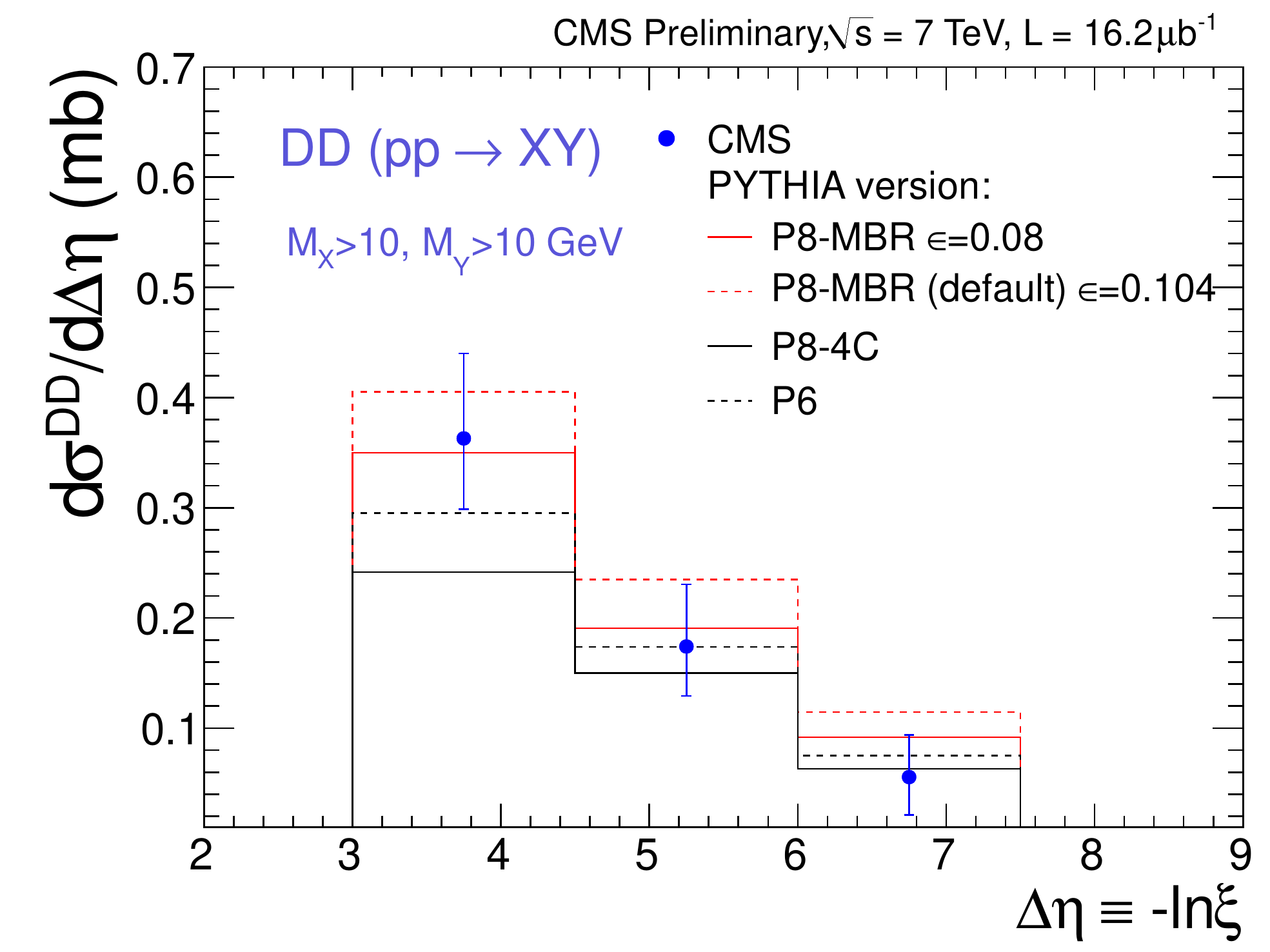}
\vspace{-0.7cm}
\caption{The SD (left) and DD (middle) cross sections as a function of $\log_{10}\xi$, and the DD cross section as a function of $\Delta\eta = -\ln\xi$ (right), compared to {\sc pythia6}, {\sc pythia8-4c} and {\sc pythia8-mbr} MC.}
\label{fig1}
\end{figure}

$~ ~\bullet$ Diffractive cross sections have been measured~\cite{FSQ-12-005} using the low-pileup 2010 data at $\sqrt{s}=7$ TeV, corresponding to an integrated luminosity of 16.2 $\mu \mbox{b}^{-1}$, and a minimum-bias sample of inelastic events, limited to using the central CMS detector ($-4.7<\eta<4.7$). Diffractive events were selected based on the presence of a forward or central Large pseudoRapidity Gap (LRG) in the CMS detetor, of at least 4 and 3 units in pseudorapidity, respectively. The forward-LRG sample consisted of approximately equal numbers of single-diffractive ($pp\rightarrow Xp$, SD) events and double-diffractive ($pp\rightarrow XY$, DD) events, for which one of the dissociated masses was low and escaped detection in the central detector. Subsamples enhanced in SD and DD events were selected by requiring an absence or a presence of an energy deposit in the CASTOR calorimeter, which covers the very forward region of the experiment, -6.6 $< \eta <$ -5.2 (in the direction of the forward LRG). The differential SD cross section as a function of $\xi$ (incoming-proton momentum loss), and the differential DD cross section as a function of $\xi_{X}=M^2_X/s$ for $0.5<log_{10}(M_Y/GeV)<1.1$ (CASTOR acceptance), after subtracting the background contribution to the signal (DD to SD and ND to DD), are shown in Figs.~\ref{fig1} (left) and (middle), respectively. Results are compared to MC models. The predictions of {\sc pythia8-mbr}~\cite{mbrnote} are shown for two values of the $\epsilon$ parameter of the Pomeron trajectory ($\alpha(t)=1+\epsilon+\alpha't$), $\epsilon = 0.08$ and $\epsilon = 0.104$. Both values describe the measured SD cross section within uncertainties, while the DD data favor the smaller value of $\epsilon$. The predictions of {\sc pythia8-4c} and {\sc pythia6} describe well the measured DD cross section, but fail to describe the falling behavior of the SD data. 
The central-LRG sample, dominated by DD events, was used to extract the differential DD cross section as a function of the central-gap width, $\Delta\eta$, for $\Delta\eta>3$, $M_X>10$ GeV and $M_Y>10$ GeV. The cross section is presented in Fig.~\ref{fig1} (right), which shows a reasonable agreement with MC predictions. 

\begin{floatingfigure}[r]{0.66\textwidth}
\vspace{-0.5cm}
\begin{center}
\includegraphics[height=2in]{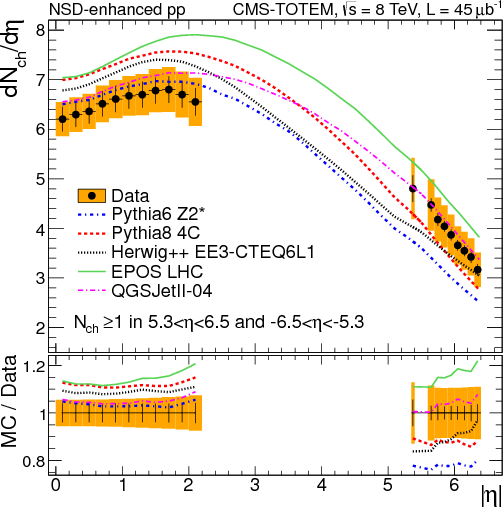}
\includegraphics[height=2in]{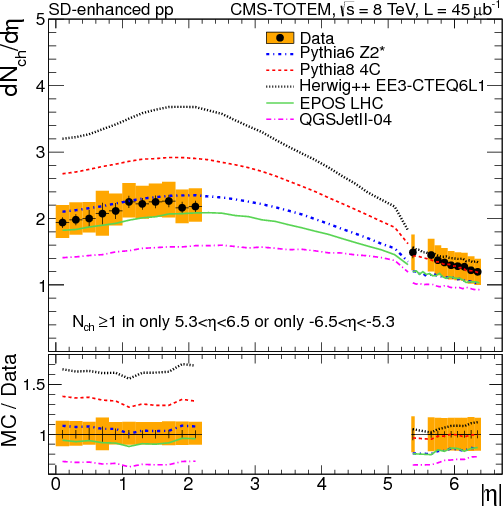}
\end{center}
\vspace{-0.6cm}
\caption{Pseudorapidity distributions, $dN_\mathrm{ch}/d\eta$, for the  NSD-enhanced (left), and SD-enhanced (right) samples, compared to MC predictions.}
\label{fig2}
\end{floatingfigure}

$\bullet$ The CMS and TOTEM collaborations have measured~\cite{FSQ-12-026} pseudorapidity distributions of charged particles, $dN_\mathrm{ch}/d\eta$, using the low-pileup 2012 data at $\sqrt{s}= 8$ TeV ($L=45~ \mu \mbox{b}^{-1}$), recorded during the common CMS and TOTEM runs with a non standard ($\beta^{*}= 90$ m) LHC optics configuration. This is the first result of the combined CMS and TOTEM analysis, covering the ranges $|\eta|<2.2$ and $5.3<|\eta|<6.4$, respectively. The minimum-bias trigger was provided by the TOTEM T2 telescopes ($5.3 <|\eta|< 6.4$). Depending on the signal configuration in the T2 detectors, events were categorized into three different samples: (i) an inclusive sample, sensitive to 91--96\% of the total inelastic proton-proton cross section, (ii) a sample enhanced in non-single diffractive (NSD-enhanced) events, and (iii) a sample enhanced in single-diffractive (SD-enhanced) events. The measured $dN_\mathrm{ch}/d\eta$ distributions for the NSD-enhanced and SD-enhanced samples are presented in Figs.~\ref{fig2} (left) and (right), respectively, showing that the charged particle density decreases with $|\eta|$. The results are compared to the predictions of various Monte Carlo models: {\sc pythia6-Z2*}, {\sc pythia8-4c}, {\sc herwig++}, {\sc epos}, and {\sc qgsjet-II-04}. None of the models provides a consistent description of the measured distributions. 

\begin{floatingfigure}[r]{0.65\textwidth}
\vspace{-0.2cm}
\begin{center}
\includegraphics[height=2.5in]{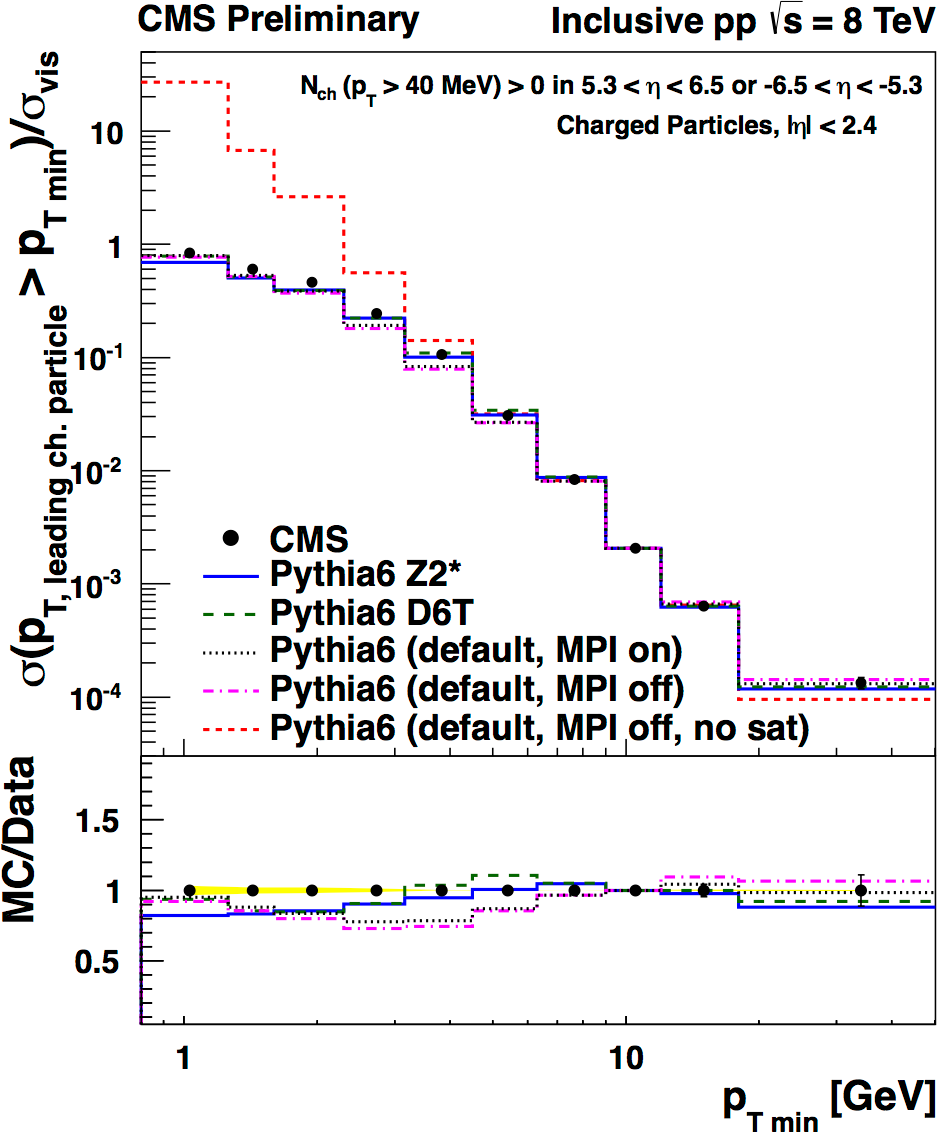}
\includegraphics[height=2.5in]{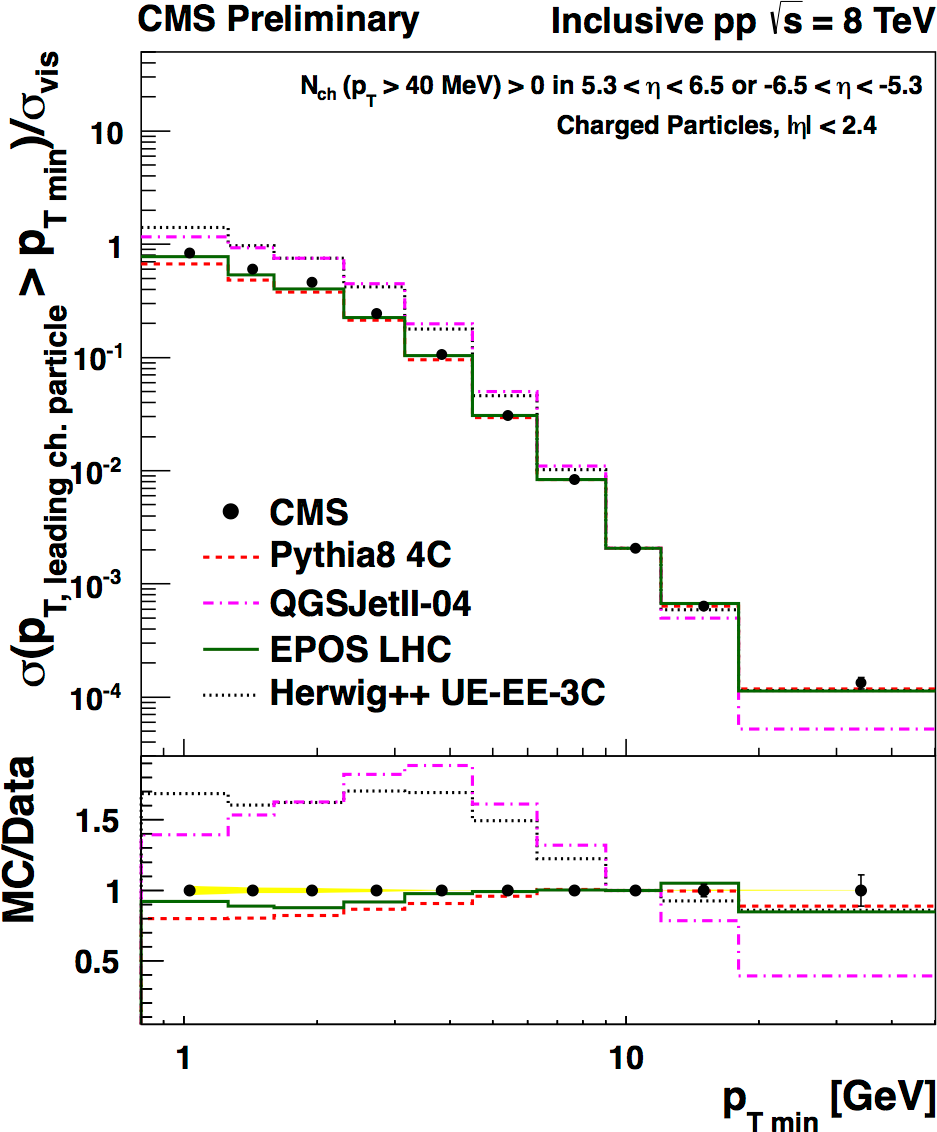}
\end{center}
\vspace{-0.7cm}
\caption{Integrated cross-section as a function of $p_\mathrm{T,min}$, compared to predictions of {\sc pythia6} (left), and various MC models (right).}
\label{fig3}
\end{floatingfigure}

$\bullet$ The integrated leading charged-particle and leading charged-particle-jet cross sections have been measured~\cite{FSQ-12-032} at $\sqrt{s}$ = 8 TeV using $45~\mu \mbox{b}^{-1}$ of the low-pileup data collected in 2012. The charged particles (charged-particle jets) were measured in the pseudorapidity range $|\eta|< 2.4 (1.9)$ for transverse momenta $p_\mathrm{T}> 0.8 (1.0)$ GeV. The measured yields integrated above a given minimum transverse momentum $p_\mathrm{T,min}$ provide information on the transition from the perturbative to the non- perturbative region, e.g. the mechanism by which the underlying parton-parton cross sections are regularized when approaching low-$p_\mathrm{T}$ non-perturbative domain. The measured integrated cross sections as a function of $p_\mathrm{T,min}$ are shown for leading charged particles in Fig.~\ref{fig3}. Figure~\ref{fig3} (left) shows the results compared to predictions of {\sc pythia6} tune {\sc Z2*} and {\sc D6T}, as well as {\sc pythia6} default with and without the simulation of multiparton interactions (MPI). Also shown is the impact of turning off the regularization of the cross section completely (``default, MPI off, no sat''). In Fig.\ref{fig3} (right) the comparison to {\sc pythia8-4c}, {\sc herwig++}, {\sc epos}, and {\sc qgsjet-II-04} is shown. The approach implemented in PYTHIA and HERWIG describes the general trend of the measurements but fails to describe the details of the transition. The prediction of EPOS agrees much better with the measurements. 

$\bullet$ The measurement of inclusive low-$p_\mathrm{T}$ jet cross section at $\sqrt{s}$ = 8 TeV has been performed~\cite{FSQ-12-031} with 5.8 pb$^{-1}$ of the data recorded during low-pileup 2012 runs. Jets were reconstructed in the rapidity region $|y|<4.7$ for jet transverse momenta $21<p_\mathrm{T}<74$ GeV. The results extend the CMS measurement of inclusive jet cross sections for jets with $p_\mathrm{T}>74$ GeV~\cite{SMP-12-012} to the region of low $p_\mathrm{T}$ and high $y$, sensitive to low-$x$ parton effects. The measured cross sections were compared (not shown) to predictions of 
NLO QCD calculations performed using {\sc NLOJET++}, which describe the data well in a wide range of $p_\mathrm{T}$ and $y$.

\begin{floatingfigure}[r]{0.38\textwidth}
\centering
\includegraphics[height=1.8in]{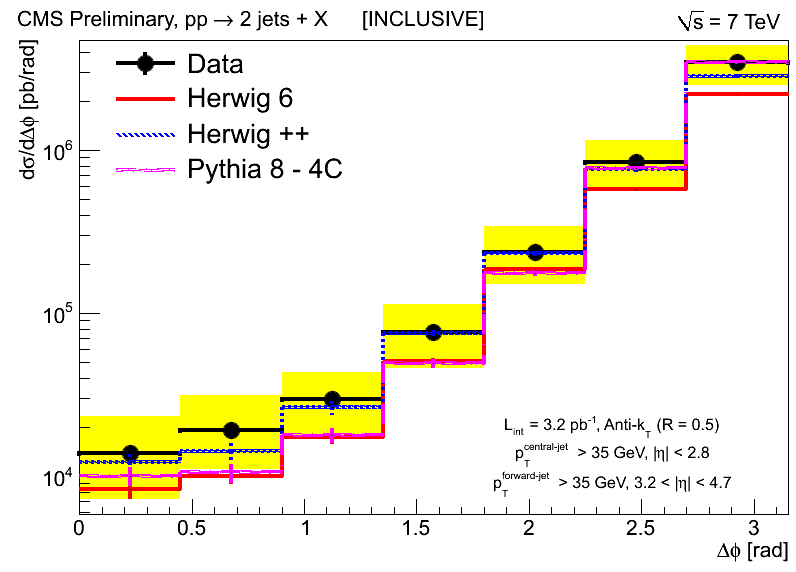}
\vspace{-0.7cm}
\caption{Cross section as a function of the azimuthal angle $\Delta \phi_{dijet}$ between the foward and the central jet, compared to predictions of various MC generators.}
\label{fig4}
\end{floatingfigure}

$\bullet$ The azimuthal angle decorrelation between forward and central jets has been studied~\cite{FSQ-12-008} at $\sqrt{s}$ = 7 TeV using 3.2 pb$^{-1}$ of the 2010 data. Events were selected containing at least one forward and one central jet with $p_\mathrm{T} > 35$ GeV. The pseudorapidities of the central and the forward jet axis were required to be $|\eta|< 2.8$ and $3.2<|\eta|< 4.7$, respectively. In lowest order QCD (Born level), two jets are produced back to back in the transverse plane of the event, and the azimuthal angular separation $\Delta \phi_{dijet} = |\phi_{jet}-\phi_{jet2}|$ is equal to $\pi$. Additional radiation leads to decorrelations, i.e. to deviations of $\Delta \phi_{dijet}$ to smaller values. Soft radiation will create only small deviations in the correlation of the two highest-$p_\mathrm{T}$ jets, while additional hard jets in the final state will lead to significant changes with respect to the dijet topology. Dijet azimuthal decorrelations are therefore a valuable tool for studying QCD radiation effects. Due to the large separation in pseudorapidity, $\Delta \eta$, the sensitivity to the details of the parton emissions along the parton ladder is maximised. The measurement is done inclusively and differentially for different $\Delta \eta$, between the jets, with the largest separation being $\Delta \eta = 7.5$. Figure~\ref{fig4} shows the inclusive cross section measured in bins of the azimuthal angle $\Delta \phi_{dijet}$ between the forward and the central jet, compared to the predictions of {\sc herwig6}, {\sc herwig++}, and {\sc pythia8-4C} simulations. The MC generators in general describe the measurement within the systematic uncertainties. When the contribution from MPI is turned off in {\sc pythia6} (not shown) the description of the dijet correlation is worse. {\sc herwig++} provides by far the best predictions describing both the shape and normalization of the data. The analysis was also carried out for two subsamples, one where an additional jet is required between the forward and the central jet, and one where the additional jet is vetoed. Similar conclusions were drawn from the comparison to the MC models.

\begin{floatingfigure}[r]{0.66\textwidth}
\vspace{-0.5cm}
\begin{center}
\includegraphics[height=2.0in]{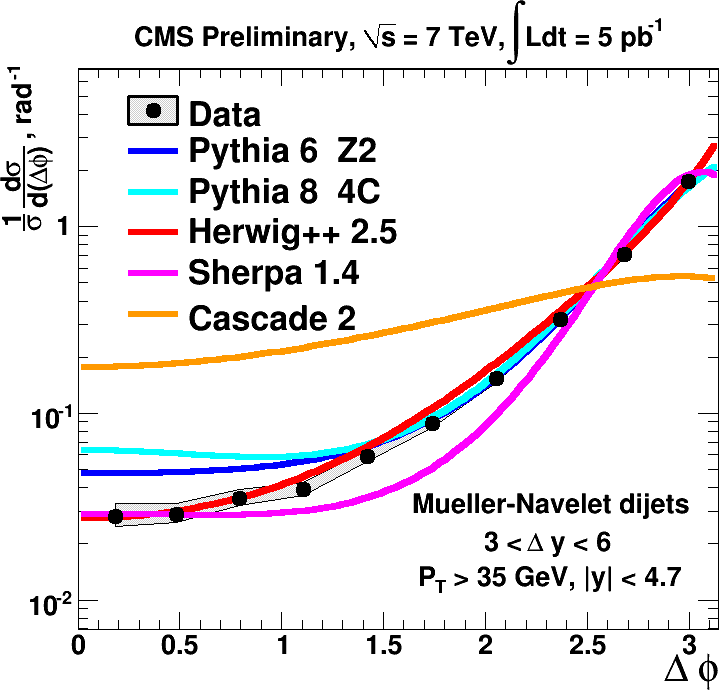}
\includegraphics[height=2.0in]{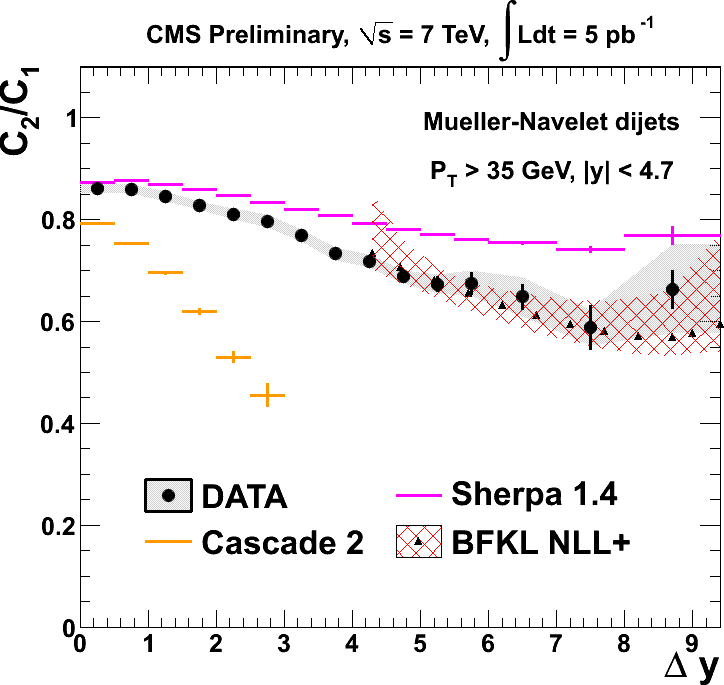}
\end{center}
\vspace{-0.6cm}
\caption{The azimuthal angle $\Delta \phi_{dijet}$ between MN jets in the rapidity interval $3<\Delta y< 6$ (left), and the ratio of average cosines $C_2/C_1$ in bins of  $\Delta y$ (right), compared to various MC models.}
\label{fig5}
\end{floatingfigure}

$\bullet$ The azimuthal angle decorrelations have also been studied~\cite{FSQ-12-002} in events with jets most forward and backward in rapidity (Mueller-Navelet, MN dijets), using 5 pb$^{-1}$ of the 2010 data at $\sqrt{s}$ = 7 TeV. Jets with $p_\mathrm{T} > 35$ GeV and $|y|$ 4.7 were considered. The requirement of two jets with similar $p_\mathrm{T}$ suppresses contributions in the\\ DGLAP scheme, which is based on $p_\mathrm{T}$ ordering. The BFKL equation predicts a strong rise of additional parton radiation with an increase of rapidity span of the\\ event, thus the study of azimuthal decorrelation of dijets as a function of rapidity separation may reveal effects beyond the DGLAP description. The measurements of the azimuthal decorrelation angle, $\Delta \phi_{dijet}$, its average cosines, $C_n = \langle\cos(n(\pi- \phi_{dijet})) \rangle$, and their ratios have been performed in bins of rapidity separation, $\Delta y$, between the jets, reaching up to $\Delta y$=9.4 for the first time. Figure~\ref{fig5} (left) shows the normalized $\Delta \phi_{dijet}$ distribution for $3<\Delta y< 6$, compared to various DGLAP based MC generators. While {\sc pythia6-z2} and {\sc pythia8-4C} predict too much decorrelation, and {\sc sherpa 1.4} too little, {\sc herwig++} provides an overall satisfactory description of the data. The result is also compared to the LL BFKL-inspired generator {\sc cascade 2}, which predicts far too much decorrelation. The ratios of cosines are expected to be more sensitive to BFKL effects, because of cancellation of DGLAP contributions. The measured $C_2/C_1$ ratio is presented in Fig.~\ref{fig5} (right), showing that the NLL BFKL calculations describe the data well; the DGLAP based predictions are not able to describe the $C_2/C_1$ distribution (not shown). 

\begin{floatingfigure}[r]{0.51\textwidth}
\vspace{-0.4cm}
\begin{center}
\includegraphics[height=2.5in]{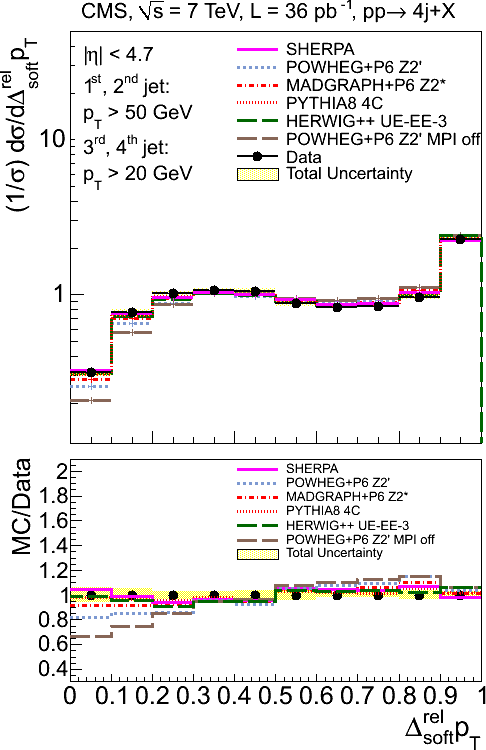}
\includegraphics[height=2.5in]{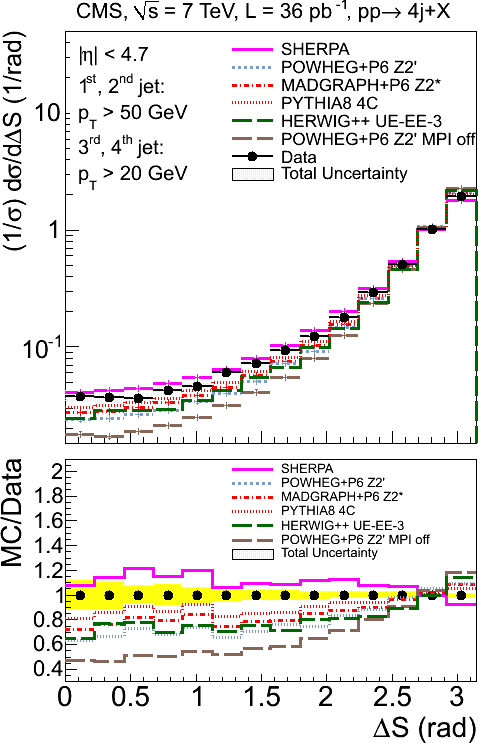}
\end{center}
\vspace{-0.7cm}
\caption{Distributions of $\Delta^\mathrm{rel}_\mathrm{soft}p_\mathrm{T}$ (left) and $\Delta S$ (right), compared to various MC predictions.}
\label{fig6}
\end{floatingfigure}

$\bullet$ The production of exactly four jets, with a pair of hard jets with $p_\mathrm{T} > 50$ GeV and another pair with $p_\mathrm{T} > 20$ GeV within $|\eta|< 4.7$, have been studied~\cite{FSQ-12-013} using the data sample at $\sqrt{s}$ = 7 TeV, collected in 2010 ($L=36~ \mu \mbox{b}^{-1}$). The integrated cross section has been measured to be $\sigma(\mbox{pp}\rightarrow \mbox{4j+X})= 330 \pm 5 (\mbox{stat}) \pm 45(\mbox{syst})$ nb. The differential cross sections have been measured as a function of $p_\mathrm{T}$ and $\eta$ for each jet, and compared to the predictions of {\sc sherpa 1.4}, {\sc pomheg}, {\sc madgraph}, {\sc pythia8-4C}, and {\sc herwig++} (not shown). The models are able to describe the differential cross sections only in some regions of the phase space. The predictions of the differential cross sections at large $p_\mathrm{T}$ are reasonable, but significant differences arise at smaller $p_\mathrm{T}$ especially for the subleading and soft jets. 

A study of correlations between jets can provide additional information on the underlying production process, and help disentangle contributions from a single parton scattering (SPS, a $2 \rightarrow 4$ partonic process) and a double parton scattering (DPS, two $2 \rightarrow 2$ processes). Cross sections measured as function of the correlation variables, defined as the balance in $p_\mathrm{T}$ between the two soft jets, $\Delta^\mathrm{rel}_\mathrm{soft}p_\mathrm{T}$, and the azimuthal angle between the hard and the soft dijet pairs, $\Delta S$, are compared  in Fig.~\ref{fig6} to the predictions of the MC simulations mentioned above. None of the MC models is able to describe the data in the regions dominated by the contribution from DPS (low $\Delta^\mathrm{rel}_\mathrm{soft}p_\mathrm{T}$ and flat $\Delta S$). This can be taken as an indication for the need of DPS in the simulations, and the data can be used to constrain MC tunes; e.g. new {\sc pythia8} tunes have recently been performed by CMS~\cite{GEN-14-001}, and when the four-jet correlation variables are included in the tuning procedure, the effective DPS cross section is predicted to be $\sigma_\mathrm{eff}=21.3^{+1.2}_{-1.6}$ mb, compared to $\sigma_\mathrm{eff}=30.3$ mb obtained for the default {\sc pythia8-4c} tune. 

\begin{floatingfigure}[r]{0.32\textwidth}
\vspace{-0.25cm}
\begin{center}
\includegraphics[height=2in]{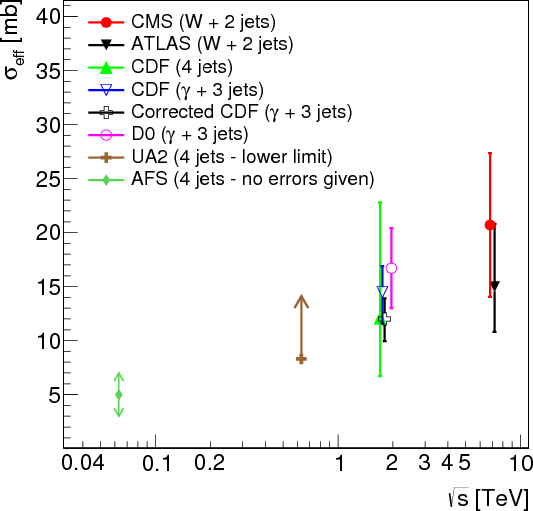}
\end{center}
\vspace{-0.75cm}
\caption{Center-of-mass energy dependence of $\sigma_\mathrm{eff}$ measured by different experiments.}
\label{fig7}
\end{floatingfigure}

$\bullet$ A study of DPS has also been performed~\cite{FSQ-12-028} with W+2-jet events, using 5 pb$^{-1}$ of the data at $\sqrt{s}$ = 7 TeV. DPS with a W+2-jet final state occurs when one hard interaction produces a W boson and another produces a dijet in the same pp collision. Events with a W boson, reconstructed from a muon of $p_\mathrm{T}> 35$ GeV and missing transverse energy of $E\!\!\!\!/_\mathrm{T} > 30$ GeV, were required to have exactly two jets with $p_\mathrm{T} > 20$ GeV and $|\eta|< 2$. The effective DPS cross section, $\sigma_\mathrm{eff}$, was measured using the relation $\sigma_\mathrm{eff}=R \cdot \sigma_\mathrm{2j}/f_\mathrm{DPS}$, where $R=N_\mathrm{W+0j}/N_\mathrm{W+2j}$ is the ratio of W+0-jet to W+2-jet events and $\sigma_\mathrm{2j}$ is the di-jet production cross section, both in the kinematic region of the W+2-jet measurement; $f_\mathrm{DPS}$ is the fraction of DPS events in the W+2-jet sample. The fraction of DPS in W+2-jet events was extracted from a DPS + SPS template fit to the distributions of the relative $p_\mathrm{T}$ balance between the two jets, $\Delta^\mathrm{rel}p_\mathrm{T}$, and of the azimuthal angle between the W boson and the dijet system, $\Delta S$. The DPS and SPS templates were obtained from the {\sc madgraph5 + pythia8} simulation, which provided a good description of the $\Delta^\mathrm{rel}p_\mathrm{T}$ and $\Delta S$ variables. The obtained value of the DPS fraction is $f_\mathrm{DPS} = 0.055 \pm 0.002~\mbox{(stat)} \pm 0.014~\mbox{(syst.)}$,  and the effective cross section is calculated to be $\sigma_\mathrm{eff} = 20.7 \pm 0.8~\mbox{(stat)} \pm 6.6~\mbox{(syst)}$ mb. The measured value of the effective cross section, shown in Fig.~\ref{fig7}, is consistent with the ATLAS and Tevatron results. A firm conclusion on the energy dependence of $\sigma_\mathrm{eff}$ cannot be drawn because of the large systematic uncertainties.

\begin{floatingfigure}[r]{0.65\textwidth}
\vspace{-0.5cm}
\begin{center}
\includegraphics[height=2in]{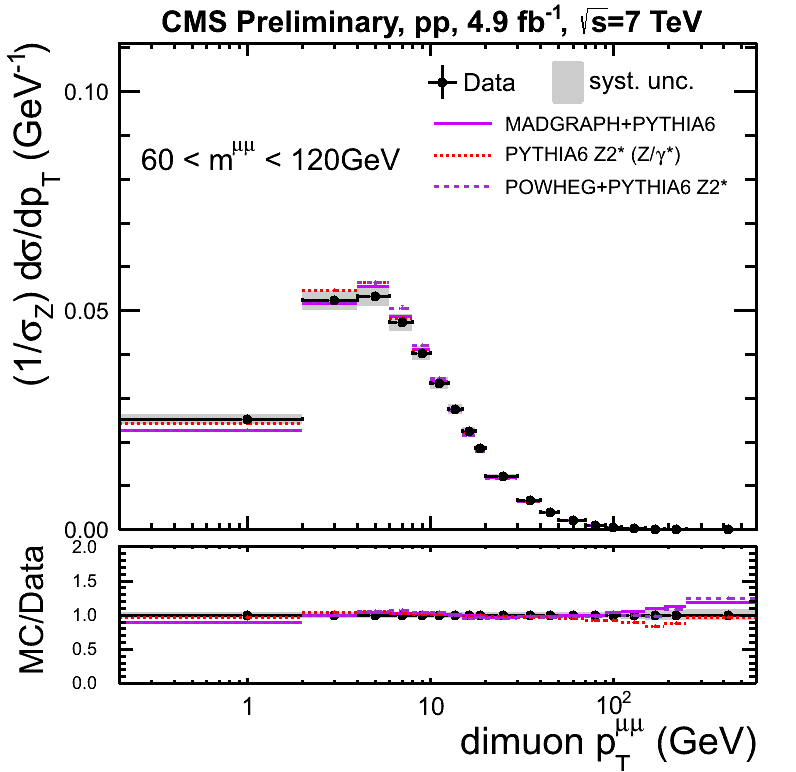}
\includegraphics[height=2in]{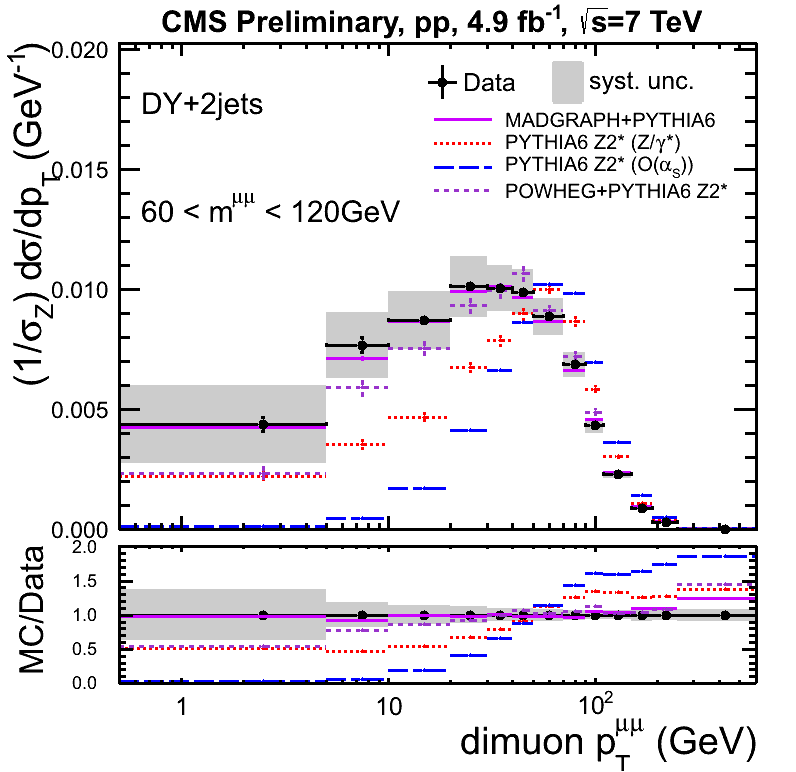}
\end{center}
\vspace{-0.7cm}
\caption{Di-muon $p^{\mu\mu}_\mathrm{T}$ spectra for the DY (left) and DY+2jets (right) processes for $60< m^{\mu\mu}<120$ GeV, compared to MC predictions.}
\label{fig8}
\end{floatingfigure}

$\bullet$  The cross sections for Drell-Yan (DY) production and DY production in association with at least one or two jets have been measured \cite{FSQ-13-003} using 4.9 pb$^{-1}$ of the data collected at $\sqrt{s}$ = 7 TeV. Differential cross sections as a function of the DY di-muon transverse momentum $p^{\mu\mu}_\mathrm{T}$ have been measured in the di-muon mass range $30 < m^{\mu\mu} < 1500$ GeV. The measurement was performed for muons with $|\eta^{\mu}|<2.1$ and $p^{\mu}_\mathrm{T}> 20~(10)$ GeV, for the leading (subleading) muon. Jets were required to have $p^\mathrm{jet}_\mathrm{T}>30$ GeV in the range $|\eta^\mathrm{jet}| < 4.5$. The di-muon $p^{\mu\mu}_\mathrm{T}$ distributions for the inclusive DY and DY + 2 jets production processes are shown in Figs.~\ref{fig8} (left) and (right), respectively. The distributions are well described by the {\sc madgraph + pythia6} simulation. For the inclusive DY case, the $p^{\mu\mu}_\mathrm{T}$ distribution is rising from small $p^{\mu\mu}_\mathrm{T}$ towards a maximum at around 5 GeV and then falling towards large $p^{\mu\mu}_\mathrm{T}$. The rising behavior at small $p^{\mu\mu}_\mathrm{T}$ is described by soft gluon resummation, and is treated in the simulation by initial state parton showers. With the DY + 2 jets selection, the range of soft gluon resummation is enlarged from about 5 GeV to about 30 GeV, thus allowing to study multiple-gluon emission and resummation effects in the perturbative region, and to separate them from the non-perturbative contribution at small $p_\mathrm{T}$. In the simulation without initial state parton shower a sharp drop of the cross section below 30 GeV is observed.

\vspace{-0.1cm}

\section{Summary}

The recent CMS results on diffraction, soft QCD and forward physics have been presented. The data can help improve pQCD-based theory predictions and the modeling of the non-perturbative part of MC simulations.

\end{document}